\documentclass[twocolumn,superscriptaddress,longbibliography,aps,pra,preprintnumbers]{revtex4-2}

\usepackage[normalem]{ulem}
\usepackage{graphicx}
\usepackage{bm}
\usepackage{color}
\usepackage{epstopdf}
\usepackage{amsmath}
\usepackage{amssymb}
\usepackage{epstopdf}
\usepackage{lipsum}

\usepackage{multirow}

\usepackage{threeparttable}

\usepackage[urlcolor=blue,colorlinks=true,citecolor=blue,linkcolor=blue,pdfstartview={FitH},bookmarks=false]{hyperref}

\usepackage{xcolor}

\graphicspath{{fig/}{./fig/}{.}}

\begin{document}

\title{
$T_{3}$Pb$_{2}${\it Ch}$_{2}$ ($T$=Pd,Pt and {\it Ch}=S,Se) with transition metal kagome net: \\
Dynamical properties, phonon nodal line, phonon surface states, and chiral phonons 
}

\author{Surajit Basak}
\email[e-mail: ]{surajit.basak@ifj.edu.pl}
\affiliation{\mbox{Institute of Nuclear Physics, Polish Academy of Sciences, W. E. Radzikowskiego 152, PL-31342 Krak\'{o}w, Poland}}

\author{Aksel Kobia\l{}ka}
\email[e-mail: ]{aksel.kobialka@unibas.ch}
\affiliation{Department of Physics, University of Basel, Klingelbergstrasse 82, CH-4056 Basel, Switzerland}

\author{Andrzej Ptok}
\email[e-mail: ]{aptok@mmj.pl}
\affiliation{\mbox{Institute of Nuclear Physics, Polish Academy of Sciences, W. E. Radzikowskiego 152, PL-31342 Krak\'{o}w, Poland}}

\date{\today}

\begin{abstract}
Shandite with Ni$_{3}$Pb$_{2}$S$_{2}$ chemical formula and R$\bar{3}$m symmetry, contains the kagome sublattice formed by the transition metal atoms.
Recent experimental results confirmed the possibility of successfully synthesizing Pd$_{3}$Pb$_{2}${\it Ch}$_{2}$ ({\it Ch}=S,Se) with the same structure.
In this paper, we theoretically investigate the dynamical properties of such compounds.
Furthermore, we study the possibility of realizing Pt$_{3}$Pb$_{2}${\it Ch}$_{2}$ with the shandite structure.
We show that the Pd$_{3}$Pb$_{2}${\it Ch}$_{2}$ and Pt$_{3}$Pb$_{2}$S$_{2}$ are stable with R$\bar{3}$m symmetry.
In the case of Pt$_{3}$Pb$_{2}$S$_{2}$, there is a soft mode, which is the source of the structural phase transition from R$\bar{3}$m to R$\bar{3}$c symmetry, related to the distortion within the kagome sublattice.
We discuss realized phonon nodal lines in the bulk phonon dispersions in upper frequency modes.
We show that the shandite structure can host the phonon surface states, with strong dependence by the surface kind.
Additionally, chiral phonons with circular motion of the Pb atoms around the equilibrium position are realized.
\end{abstract}

\maketitle

\begin{figure}[!b]
\includegraphics[width=\linewidth]{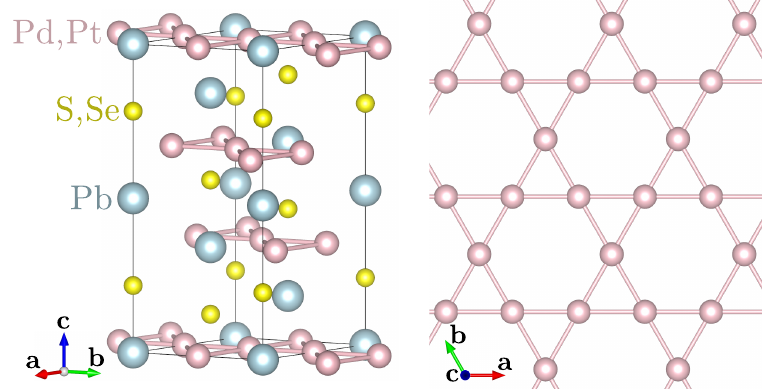}
\caption{
Conventional cell of $T_{3}$Pb$_{2}${\it Ch}$_{2}$ ($T$=Pd,Pt and {\it Ch}=S,Se) compounds with shandite structure (left panel), where the transition metal atoms form a kagome sublattice (right panel).
\label{fig.crystal}
}
\end{figure}

\section{Introduction}

The kagome lattice has the ideal electronic flat band~\cite{jiang.kang.19}.
Such features are observed in many compounds in which kagome sublattice is realized~\cite{li.zhuang.18,meier.du.20,liu.li.20,li.wang.21,nguyen.li.22,sun.zhou.22}.
Furthermore, such lattices support the emergence of the edge states, which are observed not only in solid state systems (such as multilayer silicene~\cite{li.zhuang.18}, GdV$_{6}$Sn$_{6}$~\cite{hu.wu.22} or FeGe~\cite{yin.jiang.22}), but also in acoustic~\cite{ni.gorlach.17,xue.yang.19,wu.chen.20} or photonic~\cite{zhong.wang.19,chen.lu.19,li.zhirihin.20} lattices.

In condensed matter physics, the systems possessing the kagome structure attract a lot of attention because of the flat band and edge modes, as mentioned earlier. 
In addition, such systems are characterized by a multitude of unique properties.
For example, recently discovered systems $A$V$_{3}$Sb$_{5}$ ($A$=K, Rb, Cs)~\cite{ortiz.gomes.19} with vanadium kagome lattice, are characterized by the coexistence of superconductivity and charge density waves at low temperatures~\cite{neupert.denner.22}.
Magnetic systems containing the kagome lattice~\cite{meschke.gorai.21} exhibit Weyl behavior, which has been observed, for example, in the magnetic Mn$_{3}$Sn~\cite{kuroda.tomita.17}, FeSn~\cite{kang.ye.20,xie.chen.21,kassem.tabata.16}, CoSn~\cite{meier.du.20,kassem.tabata.16,xie.chen.21,kang.fang.20} Fe$_{3}$Sn$_{2}$~\cite{ye.kang.18} and Co$_{3}$Sn$_{2}$S$_{2}$~\cite{xu.zhao.20}.
Typically, such compounds are also characterized by an intrinsic anomalous Hall effect~\cite{kida.fenner.11,wang.xu.18,liu.sun.18,thakur.vir.20,zhang.hou.22,khan.kacho.22}.

In the mentioned context, compounds with the shandite structure (Ni$_{3}$Pb$_{2}$S$_{2}$)~\cite{skinner.qian.13}, which contain the transition metal kagome sublattice, have recently become very popular.
Recently, nearly 900 compounds have been known to realize the shandite-type structure~\cite{singh.sehrawat.23}.
%%%%
Currently, probably the most popular is the topological Co$_{3}$Sn$_{2}$S$_{2}$ magnetic Weyl semimetal~\cite{kanagaraj.ning.22}.
The electronic band structure contains six nodal rings corresponding to three pairs of Weyl points in the Brillouin zone~\cite{xu.liu.18}.
This allows the emergence of the Fermi arc surface state observed experimentally~\cite{liu.liang.19}.
Moreover, the possibility of the existence of chiral edge states made this compound an excellent platform for the realization quantum anomalous Hall effect~\cite{muechler.liu.20}.

{\it Motivation.} The mineral {\it shandite}, with the chemical formula Ni$_{3}$Pb$_{2}$S$_{2}$ and R$\bar{3}$m symmetry, can be successfully synthesized under controlled conditions~\cite{skinner.qian.13}.
Substitution of Se (under S) leads to stable Ni$_{3}$Pb$_{2}$Se$_{2}$ with the same structure~\cite{range.paulus.97}.
Similarly, the substitution of Pb (under Ni) allows the realization of Pd$_{3}$Pb$_{2}$Se$_{2}$ with unchanged crystal structure~\cite{seidlmayer.bachhuber.10}, and superconductivity under pressure~\cite{yu.hua.20}.
Moreover, Pd$_{3}$Pb$_{2}$S$_{2}$ with shandite structure was reported~\cite{zabel.wandinger.14}, as well as {\it laflammeite}---the ``new'' mineral Pd$_{3}$Pb$_{2}$S$_{2}$, with C2/m symmetry was reported~\cite{barkov.martin.02}.
Theoretical study suggests that Pt$_{3}$Pb$_{2}Ch_{2}$ ({\it Ch} = S, Se) should be Dirac semi-metals with higher-order Fermi arcs~\cite{nie.chen.22}.
Lastly, Pd$_{3}$Pb$_{2}$Se$_{2}$ structure should also be stable under hydrostatic pressure~\cite{hossain.rabu.22}.
This opens a new question: can the Pt$_{3}$Pb$_{2}Ch_{2}$ ({\it Ch}=S,Se) with the shandite crystal structure be realized?
This is a very interesting problem, in the context of recent work showing the realization of Pt$_{3}$Pb$_{2}Ch_{2}$ ({\it Ch}=S,Se) with Cmcm symmetry~\cite{fang.wang.21,wang.bu.21}.
The possible formation of the structure different from shandite should not be unexpected in the context of ternary chalcogenides $M_{3}M'_{2}Ch_{2}$, for example Ni$_{3}$Bi$_{2}$S$_{2}$~\cite{sakamoto.wakeshima.06} and Rh$_{3}$Bi$_{2}$S$_{2}$~\cite{kaluarachchi.xie.15} have C2/m symmetry, while while Pd$_{3}$Bi$_{2}$S$_{2}$ possesses I2$_{1}$3 symmetry~\cite{weihrich.matar.07}.

In this manuscript, based on the {\it ab initio} technique, we show that the Pd$_{3}$Pb$_{2}${\it Ch}$_{2}$ ({\it Ch}=S,Se) and Pt$_{3}$Pb$_{2}$Se$_{2}$ are stable with shandite symmetry, while Pt$_{3}$Pb$_{2}$S$_{2}$ has imaginary phonon (soft) modes.
As a result, Pt$_{3}$Pb$_{2}$S$_{2}$  is unstable with shandite symmetry, while the displacement of the atoms induced by the soft mode leads to a distorted kagome sublattice and stable R$\bar{3}$c structure.
%%%%%%%%%%%%%%%%%%%%%%%%%%%%%%%%%%%%%%%%%
The paper is organized as follows.
Details of the computational methods are given in Sect.~\ref{sec.comp}.
Next, in Sec.~\ref{sec.res}, we present complex studies of 
$T_{3}$Pb$_{2}${\it Ch}$_{2}$ ($T$=Pd,Pt and {\it Ch}=S,Se) with shandite R$\bar{3}$m symmetry.
In particular, we discuss the crystal structure (Sec.~\ref{sec.cryst}), the main electronic properties (Sec.~\ref{sec.elec}), the basic dynamical properties (Sec.~\ref{sec.dyn}), the realized phonon nodal lines (Sec.~\ref{sec.nodal}), the phonic surface states (Sec.~\ref{sec.ph_ss}), and possible realization of the chiral phonons (Sec.~\ref{sec.chiral}).
Furthermore, we present results for novel Pt$_{3}$Pb$_{2}$S$_{2}$ with R$\bar{3}$c symmetry (Sec.~\ref{sec.pt}).
Finally, a brief summary is presented in Sec.~\ref{sec.sum}.

%Pt3Pb2S2 Cmcm (SG63)~\cite{fang.wang.21}
%Pt3Pb2Se2 Cmcm (SG63)~\cite{wang.bu.21}
%Ni$_{3}$Bi$_{2}$S$_{2}$ C2/m ~\cite{sakamoto.wakeshima.06}
%Rh$_{3}$Bi$_{2}$S$_{2}$ C2/m ~\cite{kaluarachchi.xie.15}

%%%%%%%%%%%%%%%%%%%%%%%%%%%%%%%%%%%%%%%%%%%%%
%%%%%%%%%%%%%%%%%%%%%%%%%%%%%%%%%%%%%%%%%%%%%
%%%%%%%%%%%%%%%%%%%%%%%%%%%%%%%%%%%%%%%%%%%%%

\begin{table}[!b]
\caption{
\label{tab.lattice}
Lattice parameters for optimized $T_{3}$Pb$_{2}${\it Ch}$_{2}$ compounds.
To compare, data for Ni$_{3}$Pb$_{2}$S$_{2}$ and Co$_{3}$Sn$_{2}$S$_{2}$ is also presented.
}
\begin{ruledtabular}
\begin{tabular}{lccc}
system & $a=b$~(\AA) & $c$~(\AA) & $z_\text{\it Ch}$ \\
\hline 
Pd$_{3}$Pb$_{2}$S$_{2}$ & $6.003$ & $13.675$ & $0.2174$ \\
Pd$_{3}$Pb$_{2}$Se$_{2}$ & $5.948$ & $14.483$ & $0.2104$ \\
\hline
Pt$_{3}$Pb$_{2}$S$_{2}$ & $6.216$ & $13.363$ & $0.2227$ \\
Pt$_{3}$Pb$_{2}$Se$_{2}$ & $6.124$ & $14.230$ & $0.2140$ \\
\hline
Ni$_{3}$Pb$_{2}$S$_{2}$ (Ref.~\cite{peacock.mcandrew.50}) & $5.576$ & $13.658$ & $0.285$ \\
Co$_{3}$Sn$_{2}$S$_{2}$ (Ref.~\cite{holder.dedkov.09}) & $5.375$ & $13.176$ & $0.216$
\end{tabular}
\end{ruledtabular}
\end{table}

\section{Computational details}
\label{sec.comp}

First-principles (DFT) calculations were performed using the projector augmented-wave (PAW) potentials~\cite{blochl.94} implemented in the Vienna Ab initio Simulation Package ({\sc Vasp}) code~\cite{kresse.hafner.94,kresse.furthmuller.96,kresse.joubert.99}.
Calculations are performed within the generalized gradient approximation (GGA) under the Perdew, Burke, and Ernzerhof (PBE) parameterization~\cite{perdew.burke.96}.
Wa have also included the spin--orbit coupling (SOC) and the van der Waals (vdW) corrections within the Grimme scheme (DFT-D2)~\cite{grimmie.06}.

The conventional cell of the systems (containing three formula units) were optimized using $18\times 18\times 9$ $\Gamma$-centered ${\bm k}$-grids in the Monkhorst--Pack scheme~\cite{monkhorst.pack.76}.
The calculations were performed with an energy cutoff of $350$~ eV.
The condition for breaking the optimization loop was set as the energy difference of $10^{-6}$~eV and $10^{-8}$~eV for successive steps for ionic and electronic degrees of freedom, respectively.
Symmetry of the system was analyzed by {\sc FindSym}~\cite{stokes.hatch.05} and {\sc Spglib}~\cite{togo.tanaka.18}, while momentum space analyzes were performed within {\sc SeeK-path}~\cite{hinuma.pizzi.17}.

The dynamical properties were calculated using the direct {\it Parlinski--Li--Kawazoe} method~\cite{parlinski.li.97}.
In this calculation, the interatomic force constants (IFC) are found from the force acting on the atoms.
The IFC were calculated using {\sc Phonopy} software~\cite{togo.tanaka.15} from the displacement of individual atoms.
In these calculations, the supercell containing $2 \times 2 \times 1$ conventional cells, and reduced $5 \times 5 \times 5$ $\Gamma$-centered ${\bm k}$-grids were used.
Finally, the phononic surface states were calculated using the surface Green's function technique for a semi-infinite system~\cite{sancho.sancho.85}, implemented in {\sc WannierTools}~\cite{wu.zhang.18}.

%%%%%%%%%%%%%%%%%%%%%%%%%%%%%%%%%%%%%%%%%%%%%
%%%%%%%%%%%%%%%%%%%%%%%%%%%%%%%%%%%%%%%%%%%%%
%%%%%%%%%%%%%%%%%%%%%%%%%%%%%%%%%%%%%%%%%%%%%

\section{Results and discussion}
\label{sec.res}

\subsection{Crystal structure}
\label{sec.cryst}

The $T_{3}$Pb$_{2}${\it Ch}$_{2}$ compounds crystallized typically with shandite Ni$_{3}$Pb$_{2}$S$_{2}$ structure, with R$\bar{3}$m symmetry (space group No.~166)~\cite{uaqueiro.sobany.09} (shown in Fig.~\ref{fig.crystal}).
In such a structure, the positions of the atoms are characterized by a single free parameter $z_\text{\it Ch}$, which describes the position of the chalcogenide (S or Se) atom.
The atoms occupy the respective highly symmetric Wyckoff positions: $T$ transition metal atoms occupy the $9e$ (1/2,0,0) site, Pb occupies two non-equivalent $3a$ (0,0,0) and $3b$ (0,0,1/2) positions, while {\it Ch} chalcogenide atoms occupy the $6c$ (0,0,$z_\text{\it Ch}$) site.
The transition metal atoms form the kagome sublattice, which is decorated by the Pb atom located in the same plane.

Lattice parameters of the optimized structure of the discussed compounds with shandite are collected in Tab.~\ref{tab.lattice}.
The obtained lattice parameters depend on the chemical formula of the compounds (i.e. ionic radius of substituted atoms).
Nevertheless, they are close to those of other ternary chalcogenides~\cite{singh.sehrawat.23}, such as Co$_{3}$Pb$_{2}$S$_{2}$ with $a = 5.495$~\AA\ and $c = 13.719$~\AA, or Mn$_{3}$Sb$_{2}$Te$_{2}$ with $a = 5.943$~\AA\ and $c = 14.745$~\AA.
Similarly, the free parameter $z_\text{\it Ch}$ is closer to that reported in the Co$_{3}$Sn$_{2}$S$_{2}$~\cite{holder.dedkov.09}, than in the clean shandite Ni$_{3}$Pb$_{2}$S$_{2}$~\cite{peacock.mcandrew.50}.

%lattice param. 2207.03365 table~\cite{singh.sehrawat.23}

\begin{figure*}
\includegraphics[width=\linewidth]{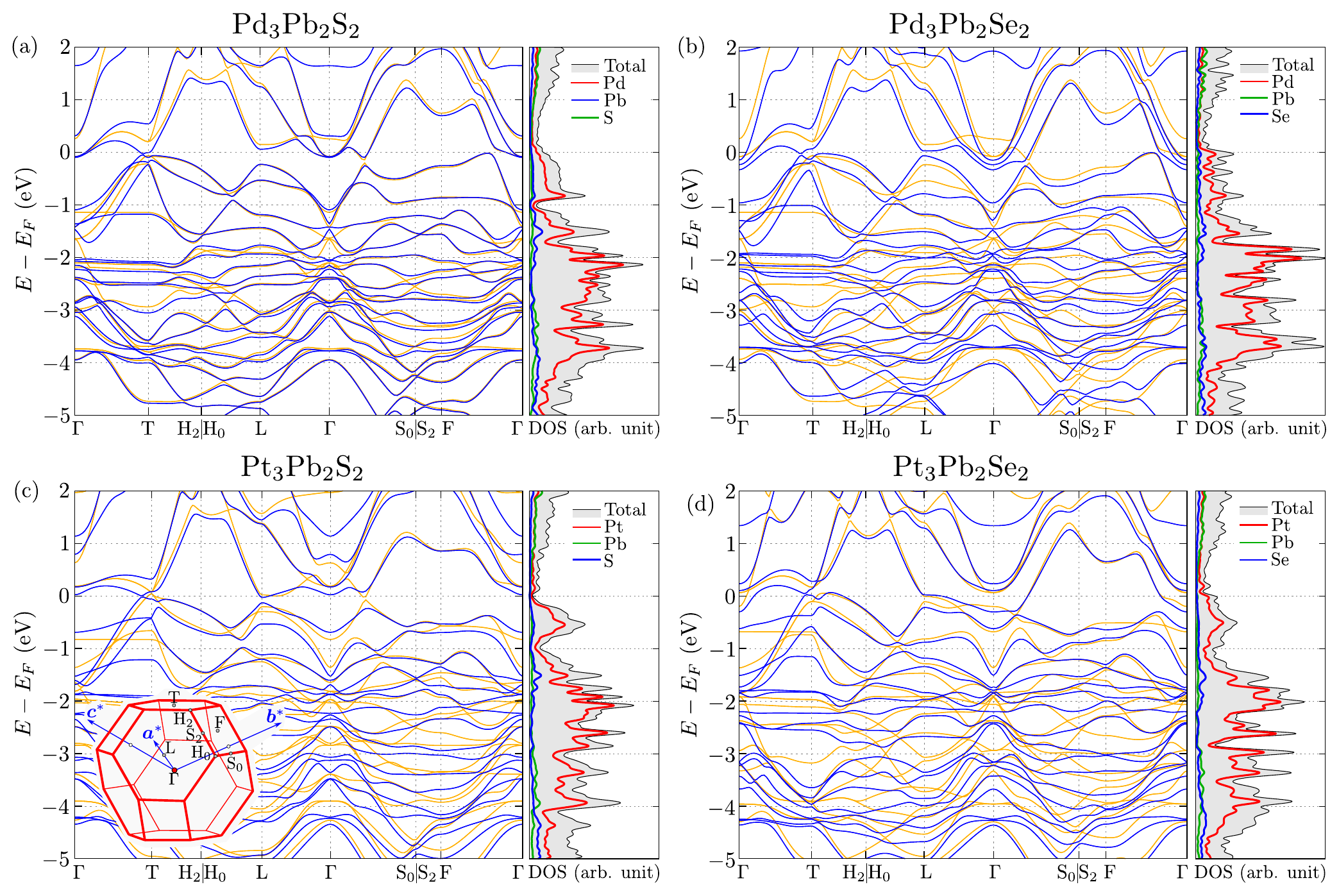}
\caption{
The electronic band structure and density of states for discussed compounds (as labeled).
The orange and blue lines on the left panels, correspond to the electronic band structure in the absence and presence of spin--orbit coupling, respectively.
Circles mark the high symmetry points, while the inset presents the Brillouin zone and its high symmetry points.
\label{fig.el_band}
}
\end{figure*}

\subsection{Electronic band structure}
\label{sec.elec}

The electronic band structure of the investigated compounds with R$\bar{3}$m shandite structure is shown in Fig.~\ref{fig.el_band}.
Theoretical results do not reveal any magnetic order in such compounds.
The electronic band structure is similar to the previously reported Pd$_{3}$Pb$_{2}${\it Ch}$_{2}$ ({\it Ch}=S,Se) by Nie {\it et al.} in Ref.~\cite{nie.chen.22}.
In each compound, we observed a strong role of the SOC on the band structure (cf. the orange and blue lines at Fig.~\ref{fig.el_band}, for results without and with SOC, respectively).
This is crucial around $\Gamma$ point, where SOC leads to a strong decoupling of the bands.
Along the $\Gamma$--T direction, the type-I Dirac point and the band inversion around the T point are realized~\cite{nie.chen.22}.
Theoretical investigations of the electronic band structure~\cite{nie.chen.22}, based on the calculation of the symmetry-based indicators $\mathrm{Z}_{2,2,2,4} = (111;0)$~\cite{fu.kane.07,po.vishwanath.17,song.zhang.18,khalaf.po.18} suggest possible electronic edge states.
However, the absence of the magnetic order does not allow the emergence of the Weyl behavior reported in Co$_{3}$Sn$_{2}$S$_{2}$~\cite{xu.zhao.20}.
However, further research on the topological properties of such compounds is necessary.

Several flat bands originating from the kagome sublayers are visible in the electronic band structures (right part of the panels in Fig.~\ref{fig.el_band}).
This is reflected in the electronic density of states (DOS), in the form of distinct peaks.
Indeed, the partial DOS analyses show that the peaks are related to the Pt atoms forming kagome net.
All peaks are well below the Fermi level, below $-1$~eV.

%%%%%%%%%%%%%%%%%%%%%%%%%%%%%%%%%%%%%%%%
%%%%%%%%%%%%%%%%%%%%%%%%%%%%%%%%%%%%%%%%
%%%%%%%%%%%%%%%%%%%%%%%%%%%%%%%%%%%%%%%%
%%%%%%%%%%%%%%%%%%%%%%%%%%%%%%%%%%%%%%%%

\begin{table}[!b]
\caption{
\label{tab.modes}
Characteristic frequencies (THz) and symmetries of the modes at the $\Gamma$ point for discuss compounds with with R$\bar{3}$m symmetry.
}
\begin{ruledtabular}
\begin{tabular}{rlrlrlrl}
\multicolumn{2}{c}{Pd$_{3}$Pb$_{2}$S$_{2}$} & \multicolumn{2}{c}{Pd$_{3}$Pb$_{2}$Se$_{2}$} & \multicolumn{2}{c}{Pt$_{3}$Pb$_{2}$S$_{2}$} & \multicolumn{2}{c}{Pt$_{3}$Pb$_{2}$Se$_{2}$} \\
\hline 
$1.23$ & ($E_\text{u}$) & $1.21$ & ($A_\text{1u}$) & $0.57$ & ($A_\text{1u}$) & $0.79$ & ($A_\text{1u}$) \\
$1.31$ & ($A_\text{1u}$) & $1.24$ & ($E_\text{u}$) & $1.32$ & ($E_\text{u}$) & $1.28$ & ($E_\text{u}$) \\
$1.53$ & ($A_\text{2u}$) & $1.50$ & ($E_\text{u}$) & $1.38$ & ($A_\text{2u}$) & $1.45$ & ($A_\text{2u}$) \\
$1.85$ & ($E_\text{u}$) & $1.66$ & ($A_\text{2u}$) & $1.87$ & ($E_\text{u}$) & $1.69$ & ($E_\text{u}$) \\
$2.35$ & ($E_\text{u}$) & $1.90$ & ($E_\text{u}$) & $2.31$ & ($E_\text{u}$) & $2.00$ & ($E_\text{u}$) \\
$2.49$ & ($A_\text{2u}$) & $2.45$ & ($A_\text{2u}$) & $2.40$ & ($A_\text{2u}$) & $2.43$ & ($A_\text{2u}$) \\
$2.85$ & ($E_\text{u}$) & $2.84$ & ($E_\text{u}$) & $2.85$ & ($E_\text{u}$) & $2.56$ & ($E_\text{u}$) \\
$3.72$ & ($A_\text{2u}$) & $3.21$ & ($A_\text{2u}$) & $3.93$ & ($A_\text{2u}$) & $3.05$ & ($A_\text{2u}$) \\
$4.81$ & ($E_\text{g}$) & $4.35$ & ($E_\text{g}$) & $5.70$ & ($E_\text{g}$) & $5.09$ & ($E_\text{g}$) \\
$5.63$ & ($E_\text{u}$) & $5.20$ & ($E_\text{u}$) & $6.59$ & ($E_\text{u}$) & $6.08$ & ($E_\text{u}$) \\
$6.38$ & ($A_\text{2u}$) & $6.27$ & ($A_\text{1g}$) & $6.84$ & ($A_\text{2u}$) & $6.98$ & ($A_\text{2u}$) \\
$6.39$ & ($A_\text{1g}$) & $6.44$ & ($A_\text{2u}$) & $7.21$ & ($A_\text{1g}$) & $7.06$ & ($A_\text{1g}$) \\
\end{tabular}
\end{ruledtabular}
\end{table}

\begin{figure*}
\includegraphics[width=\linewidth]{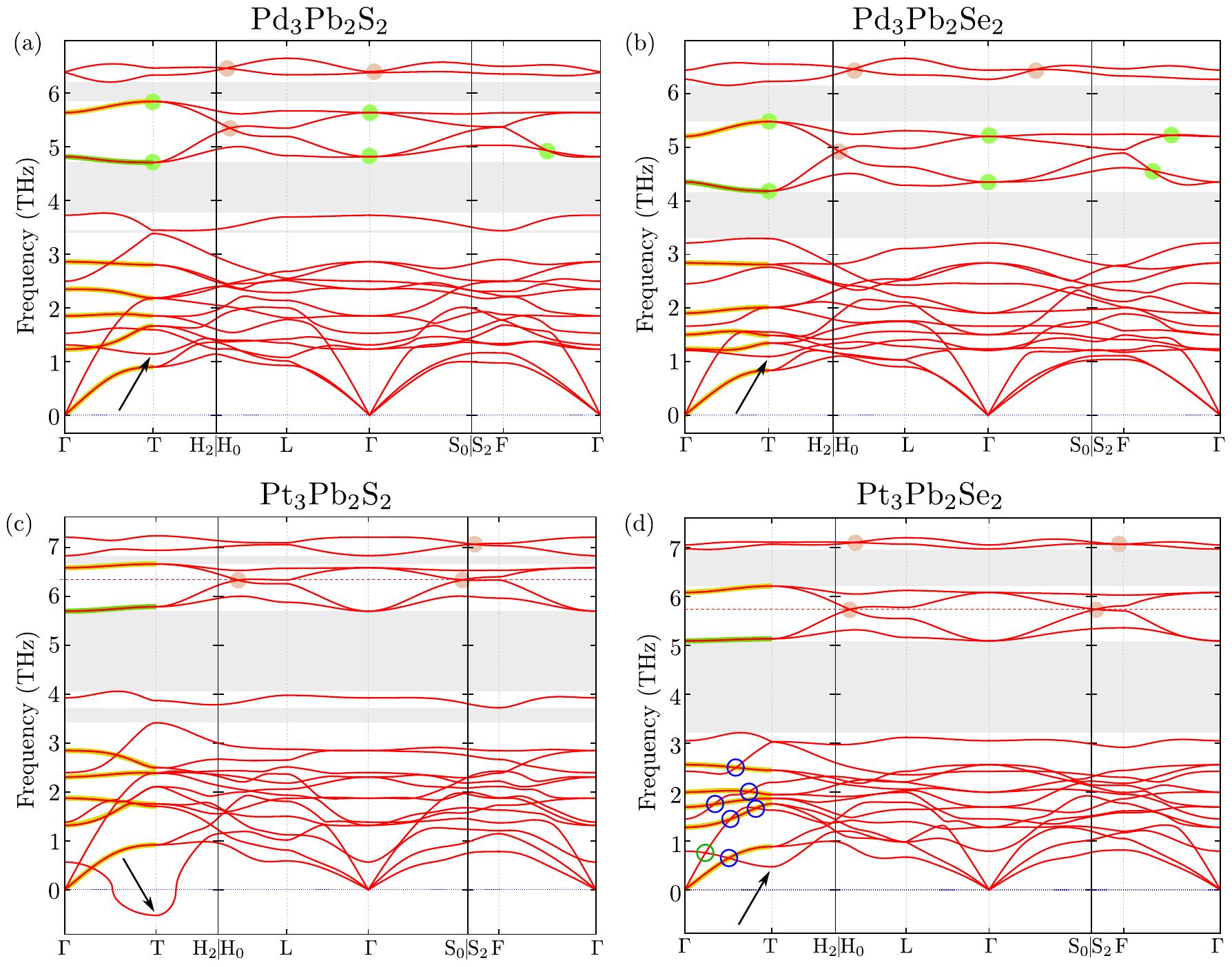}
\caption{
The phonon dispersions for the discussed compounds (as labeled).
Highlighted modes at $\Gamma$--T pat denote doubly degenerated modes with initial $E_\text{u}$ and $E_{g}$ irreducible representations at $\Gamma$ point (yellow and green solid line, respectively).
Grey areas denote the band gaps, circles denote phonon degenerate Dirac points.
\label{fig.ph_band}
}
\end{figure*}

\subsection{Dynamical stability}
\label{sec.dyn}

\begin{figure}
\includegraphics[width=\linewidth]{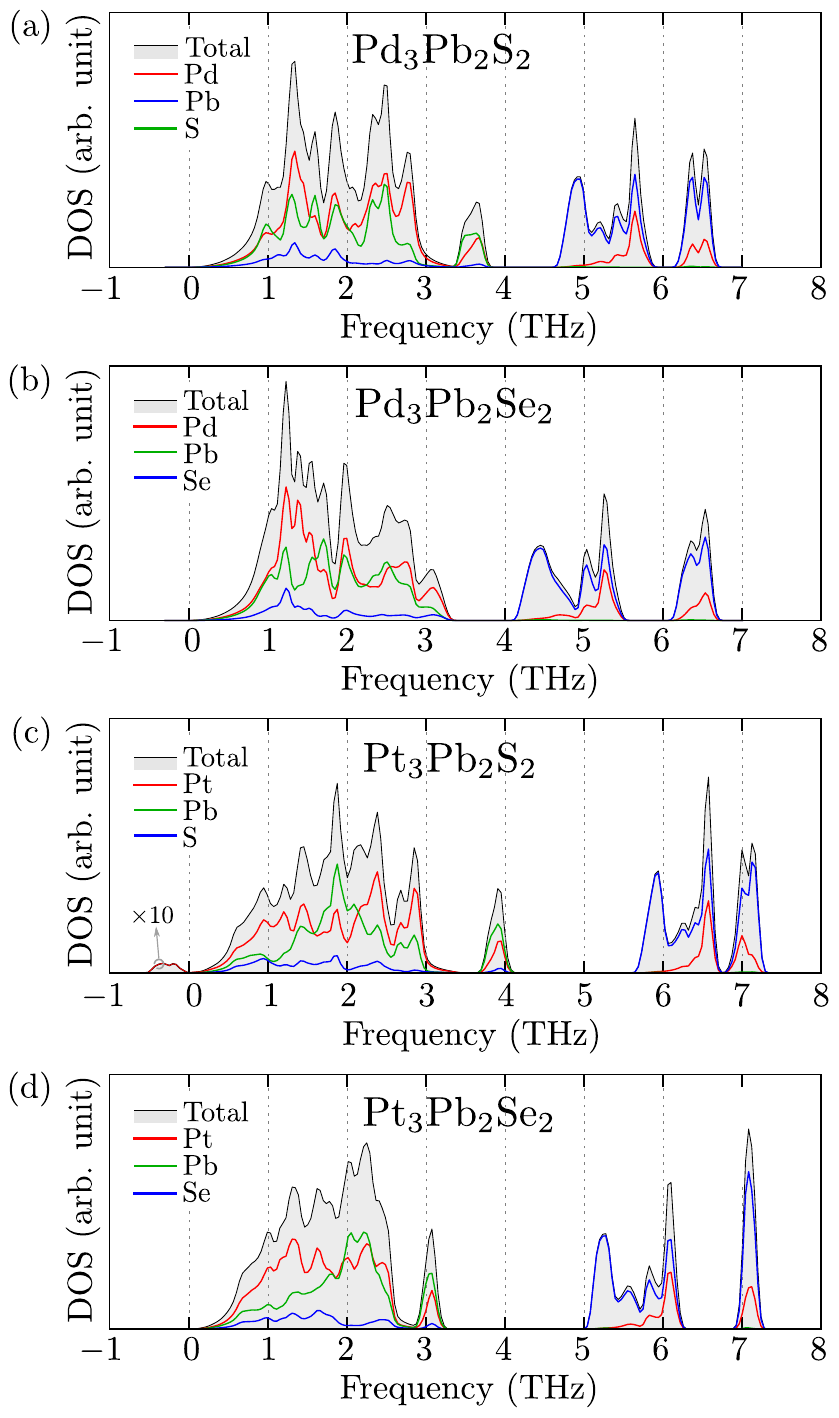}
\caption{
The phonon density of states for discuss compounds (as labeled).
For Pt$_{3}$Pb$_{2}$S$_{2}$, to guide the eye, the magnitude for softmodes was increased 10 times.
\label{fig.ph_dos}
}
\end{figure}

{\it Phonon dispersions.} The phonon dispersion curves are presented in Fig.~\ref{fig.ph_band}.
All of the discussed compounds possess phonon dispersions with similar features.
First, independently of the chemical formula, the phonon branches form several groups, separated by the band-gaps (marked by a gray area on Fig.~\ref{fig.ph_band}).
Analyses of the phonon DOS (Fig.~\ref{fig.ph_dos}) show that the branches with lowest frequencies (below $4$~THz) are related to the vibration of all modes. 
Only in this range of frequencies, Pb atoms participate in the phonon vibrations, while for higher frequencies the Pb contribution is negligible.
The modes above $4$ THz are related to the collective vibration modes of the transition metal and chalcogenide atoms.

\begin{figure*}
\includegraphics[width=\linewidth]{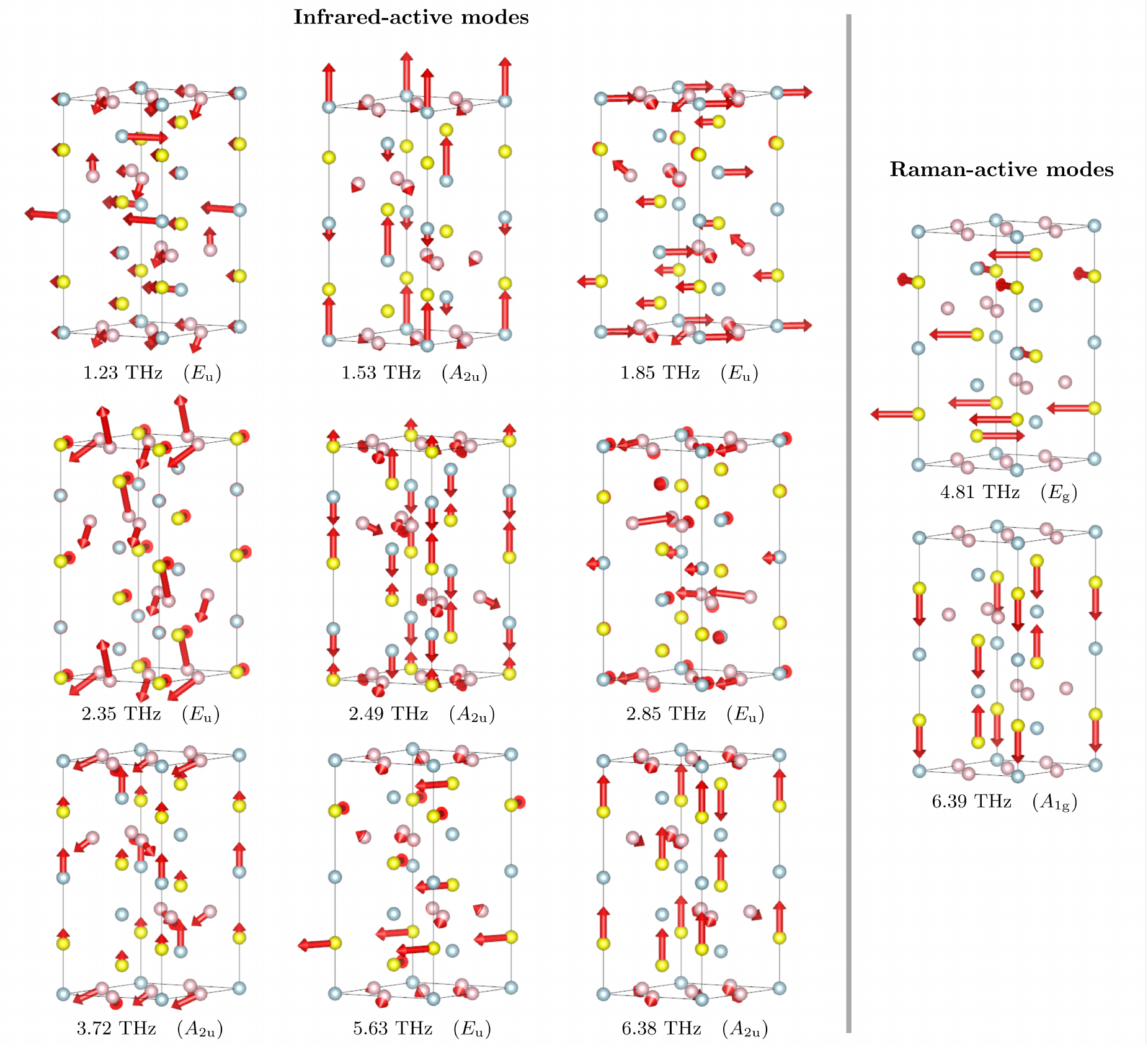}
\caption{
Schematic illustration of infrared and Raman active modes in Pd$_{3}$Pb$_{2}$S$_{2}$.
\label{fig.act_mode}
}
\end{figure*}

{\it IR and Raman active modes.}
The phonon modes at the $\Gamma$ point can be decomposed into irreducible representations of the space group R$\bar{3}$m as follows:
\begin{eqnarray}
\label{eq.irr_r3m} \Gamma_\text{acoustic} &=& A_\text{2u} + E_\text{u} , \\
\nonumber \Gamma_\text{optic} &=& A_\text{1g} + A_\text{1u} + 4 A_\text{2u} + 5 E_\text{u} + E_\text{g} .
\end{eqnarray}
In total, there are 21 vibrational modes, seven nodegenerate ($A_\text{1u}$, $A_\text{1g}$, and $A_\text{2u}$) and seven doubly degenerate ($E_\text{u}$ and $E_\text{g}$).
From this, the $A_\text{2u}$ and $E_\text{u}$ modes are infrared (IR) active, while $A_\text{1g}$ and $E_\text{g}$ are Raman active.
Selective rules for this Raman active modes were described in earlier studies of $T$Bi$_{2}$Te$_{4}$ ($T$=Mn,Fe)~\cite{kobialka.sternik.22}.
From this, the Raman modes $A_\text{1g}$ and $E_\text{g}$ can be easily distinguished during the Raman scattering measurements (with linear or circular polarized light), because of the different intensity.

The observed characteristic frequencies of the modes at $\Gamma$ points and their irreducible representations for discussed compounds with R$\bar{3}$ symmetry are collected in Tab.~\ref{tab.modes}, while the schematic illustration of the IR and Raman active modes are presented in Fig.~\ref{fig.act_mode}.
The Raman active modes are realized only by chalcogenide atoms.
In contrast to this, all atoms contribute in the IR active modes.

{\it Modes softening and system stability.}
From the comparison of the phonon dispersion curves of all compounds, we can find softening of some branches at the T point (shown by black arrows in Fig~\ref{fig.ph_band}).
In the case of Pd$_{3}$Pb$_{2}${\it Ch}$_{2}$ ({\it Ch}=S,Se) described modes at the T point are located around $1$ THz.
For Pt$_{3}$Pb$_{2}$Se$_{2}$ [Fig~\ref{fig.ph_band}(d)], this mode is clearly shifted to the lower frequencies (aorund $0.5$~ THz, below the acoustic branches).
However, the most important role is played by this mode in the case of Pt$_{3}$Pb$_{2}$S$_{2}$, where we have an imaginary frequency presented in Fig.~\ref{fig.ph_band}(c) as a negative one.
This means that only Pt$_{3}$Pb$_{2}$S$_{2}$ cannot be stable with R$\bar{3}$m symmetry. 
The exact analysis of this mode will be presented in Sec.~\ref{sec.pt}.
Here, we should also note that the softening branch is related to the lowest nondegenerate $A_\text{1u}$ mode at the $\Gamma$ point.

\begin{figure*}
\includegraphics[width=\linewidth]{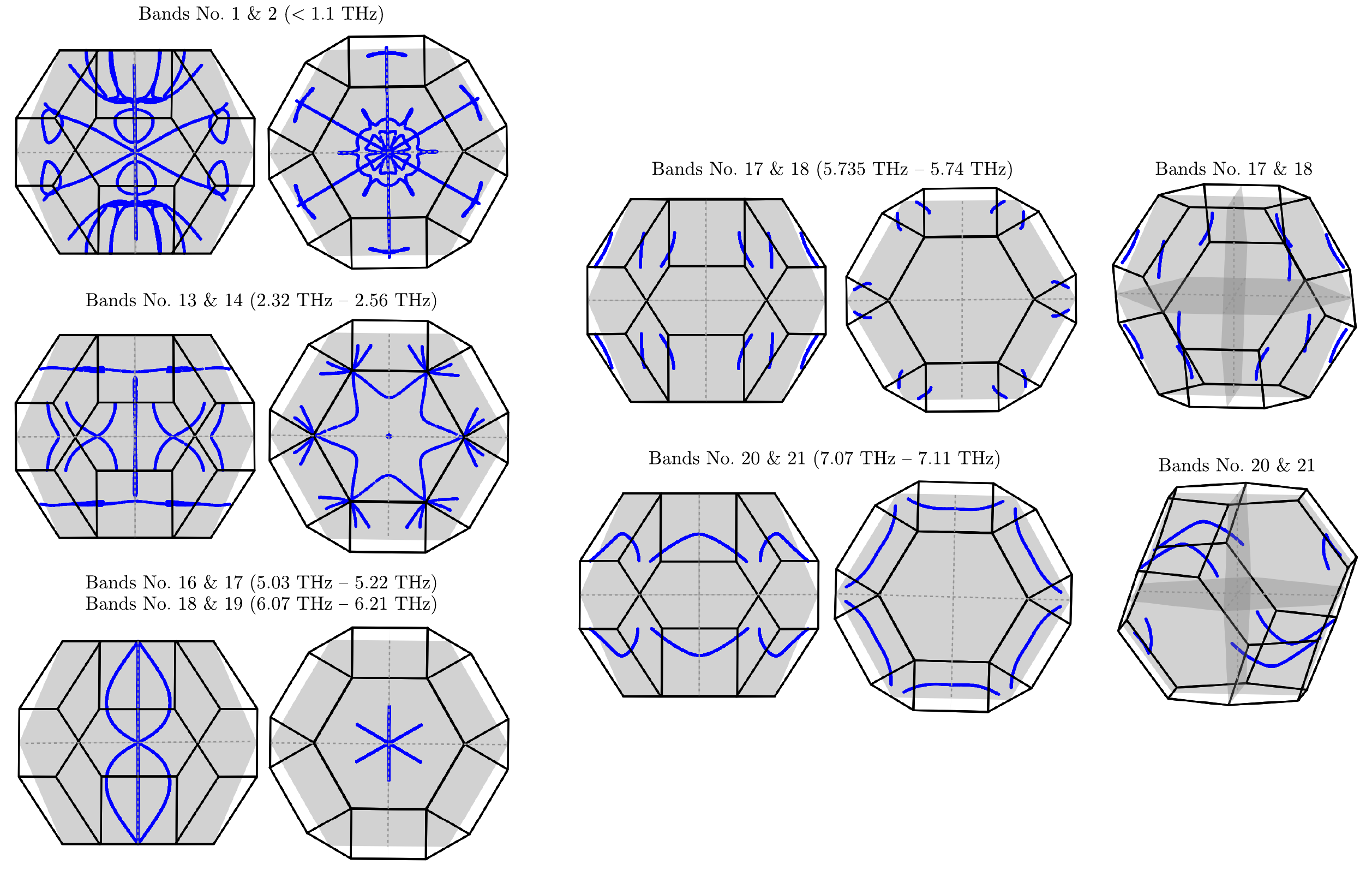}
\caption{
Example of the phonon nodal lines between different bands (as labeled) realized in Pt$_{3}$Pb$_{2}$Se$_{2}$.
Left and right panels, show top and front view on the Brillouin zone.
For nodal lines coming from bands No. 17 and 18 (20 and 21) also  a general view is presented (right column).
The nodal lines are realized in the  range of frequencies show above the Brillouin zones.
\label{fig.nodal}
}
\end{figure*}

\subsection{Phonon nodal lines}
\label{sec.nodal}

The space group R$\bar{3}$m symmetry is characterized by space inversion $\mathcal{I}$, three-fold rotation axis $3_{(111)}$, two-fold rotation axes $2_{(1\bar{1}0)}$, $2_{(10\bar{1})}$, $2_{(01\bar{1})}$, and mirror planes $\mathcal{M}_{(1\bar{1}0)}$, $\mathcal{M}_{(10\bar{1})}$, $\mathcal{M}_{(01\bar{1})}$ (in the primitive unit cell basis).
The three-fold rotation axis $3_{(111)}$ [along ${\bm c}$ direction in Fig.~\ref{fig.crystal}(a)] leads to preservation of the degeneration of the branches along the rotation axis. 
As a result, the degeneracy at the $\Gamma$ point is the same as the branches along the $\Gamma$--T direction (to guide the eye, the double generated branches are marked by yellow and gray lines in Fig.~\ref{fig.el_band}).
This leads to the emergence of typical symmetry that enforces the phonon nodal line along the $\Gamma$--T direction.

Time-reversal invaraint momentum points include one $\Gamma$, one T, three L, and three F symmetry points.
The little group analyzes are allowed for 6 irreducible representations at $\Gamma$ and T points.
This is in agreement with Eq.~(\ref{eq.irr_r3m}) describing irreducible representations at $\Gamma$ (4 possible nondegenerate and 2 possible double-degenerate representations).
In the case of L and F points, only 4 nondegenerate representations are allowed.
Due to the arguments from the previous paragraph, there are no other degenerate branches in the phonon dispersion other than the one along the $\Gamma$--T direction (according to Tab.~\ref{eq.irr_r3m}).

However, the presence of inversion symmetry and time-reversal symmetry allows for the realization of other nodal lines.
For example, in the case of MoB$_{2}$ with R$\bar{3}$m symmetry, a helical topological nodal line protected by $\mathcal{PT}$ symmetry (i.e., combination of time-reversal symmetry $\mathcal{T}$ and inversion symmetry $\mathcal{P}$) is realized by the phonon branches in ``higher'' frequency mode~\cite{zhang.miao.19}.
A similar situation is realized in our case, while the nodal line has another character.
Few examples of nodal lines realized in Pt$_{3}$Pb$_{2}$Se$_{2}$ are presented in Fig.~\ref{fig.nodal}.
In the case of the highest modes (bands No. 20 \& 21) the  nodal lines form a closed contour through the Brillouin zone in the $xy$ plane. 
Contrary to this, the nodal lines coming from crossing band No. 17 \& 18 form open lines mostly along the $z $ direction.
In both cases, the nodal lines are related to mostly constant frequencies --- in range 7.07 THz -- 7.11 THz for No. 20 \& 21, and in range 5.73 THz -- 5.74 Thz for No. 17 \& 18.
Both band crossing points are marked by green and red filled circles in Fig.~\ref{fig.ph_band}.
The degeneracy of the bands along the $\Gamma$ --T direction is reflected in the shape of the nodal lines  between bands No. 16 \& 17, and No. 18 \& 19.
Degenerate bands form the nodal lines along this direction (vertical line along the Brillouin zone).
However, additional bands crossing out of the $\Gamma$--T direction create a star-like shape of the nodal line (visible on top of view panel).
Two fold rotational symmetries lead to the additional rotation of these nodal lines in the lower half of the Brillouin zone, with respect to the upper half.
Finally, the total shape of the nodal lines takes the form of a shifted hourglass.
In these cases, nodal lines are realized for frequencies 6.07 THz -- 6.21 THz for band No. 18 \& 19, and 5.03 THz -- 5.22 THz for band No. 16 \& 17. 
Similarly to the case of MoB$_{2}$ with R$\bar{3}$m symmetry~\cite{zhang.miao.19}, the described nodal lines are protected by $\mathcal{PT}$ symmetry.

The nodal lines can also be found in the lower frequencies range.
Here, the situation is more complicated.
In the case of lower frequencies, many crossings of the non-degenerate and double-degenerate branches can be found [few examples are marked by blue circles in Fig.~\ref{fig.ph_band}(d)], which allow the realization of higher degenerate points.
For example, the nodal line between band No. 1 \& 2 is in range of from 0 THz to 1.1 THz, and is related to very complex structure containing lines and rings.
However, comparison of phonon dispersion curves for all the discussed compounds with R$\bar{3}$m symmetry (Fig.~\ref{fig.ph_band}) suggests an incidental source of these nodal lines.
On the other hand, this opens up a new possibility for nodal lines engineering via atom substitution~\cite{li.ren.12}.

From the experimental point of view, the realization of the described nodal lines can be confirmed using meV-resolution inelastic x-ray scattering (IXS)~\cite{zhang.miao.19}

%%%%%%%%%%%%%%%%%%%%%%%%%%%%%%%%%%%%%%%%%%%%%%
%%%%%%%%%%%%%%%%%%%%%%%%%%%%%%%%%%%%%%%%%%%%%%
%%%%%%%%%%%%%%%%%%%%%%%%%%%%%%%%%%%%%%%%%%%%%%

\begin{figure*}
\includegraphics[width=\linewidth]{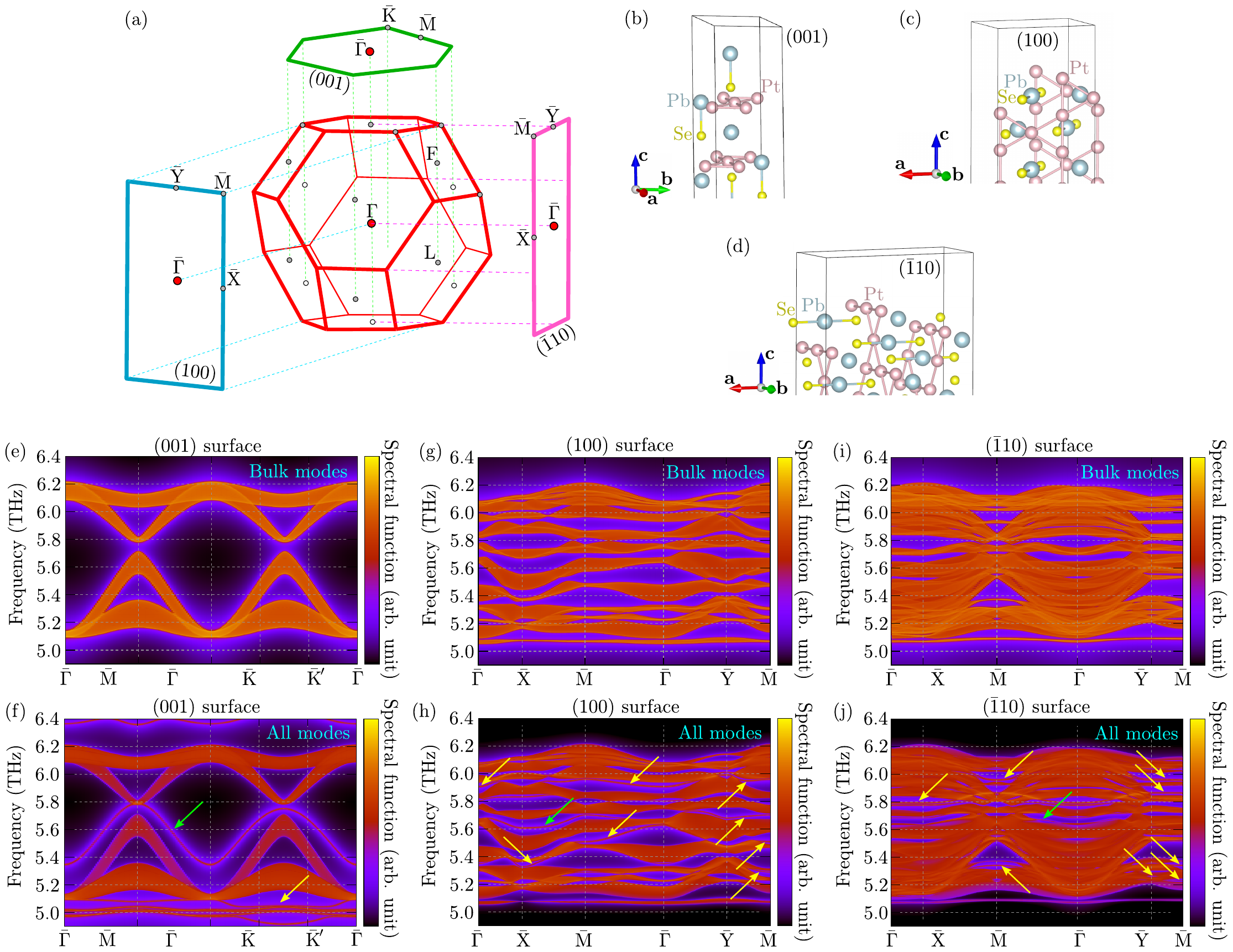}
\caption{
(a) Projection of the three dimensional bulk Brillouin zone into the two dimensional surface Brillouin zone (for different surfaces, as labeled).
(b)--(d) Termination for discussed surfaces (as labeled).
(e)--(f) Spectral function for different surfaces (as labeled). 
To compare the above, we present bulk state spectra (upper panels) and spectra containing all modes (lower panels).
Yellow and green arrows indicate surface states.
\label{fig.ph_ss}
}
\end{figure*}

\subsection{Phononic surface states}
\label{sec.ph_ss}

In this section, we focus on the phonon surface states in the high-frequency range.
We discuss here the surface states for Pt$_{3}$Pb$_{2}$Se$_{2}$, however, conclusions should be similar to those for other compounds as well.
In our investigation, we calculated the phonon spectral function using the Green function technique for different surfaces presented in Fig.~\ref{fig.ph_ss}.
Additionally, in this range of frequencies, the (bulk) phonon modes are related to the Pt and Se atom vibrations [see Fig.~\ref{fig.ph_dos}(d)].
This allows for the study of the surface modes associated with the kagome lattice of the transition metal atoms, i.e. Pt in our case. 
Here, we chose a few different surfaces realizing three types of terminations:
%%%%%%%%%%%%%%%%
\begin{itemize}
\item surface (001) in the conventional cell basis [surface (111) in the primitive unit cell basis] --- realizes the surface perpendicular to the kagome net [Fig.~\ref{fig.ph_ss}(b), with normal vector parallel to ${\bm c}$ in Fig.~\ref{fig.crystal}].
%%%%%%%%%%%%%%%%
\item surface (100) in the conventional cell basis [surface ($\bar{1}$01) in the primitive unit cell basis] --- realizes the surface perpendicular to the kagome net terminated by the triangular lattice [Fig.~\ref{fig.ph_ss}(c), with normal vector parallel to ${\bm a}$ in Fig.~\ref{fig.crystal}].
%%%%%%%%%%%%%%%%
\item surface ($\bar{1}$10) in the conventional cell basis [surface (2$\bar{1}\bar{1}$) in the primitive unit cell basis] --- realizes the surface parallel to the kagome net terminated by the chain of Pt atoms [Fig.~\ref{fig.ph_ss}(d), with normal vector parallel to ${\bm b}-{\bm a}$ in Fig.~\ref{fig.crystal}].
\end{itemize}
%%%%%%%%%%%%%%%%

The spectra for different surfaces are presented in Fig.~\ref{fig.ph_ss}(e)--(j).
As we can see, the spectra are more complex than those observed in monoatomic systems such as graphene~\cite{li.wang.20}.
To extract the surface states more straightforwardly, we present spectra of bulk modes (upper panels), and spectra containing all modes (lower panels). 
In the case of (001) surface [Fig.~\ref{fig.ph_ss}(b), (e) and (f)] there is one edge mode in the range of bulk states (green arrow).
However, out of this range, several groups of edge modes are well visible (yellow arrow).
This is true for frequencies below 5.1~THz, as weel above 6.3~THz. 
These modes can be related to the vibrations of the ``free standing'' Pt--Se molecule at the (001) surface termination [Fig.~\ref{fig.ph_ss}(b)].
In the case of (100) and ($\bar{1}$10) surfaces, the spectra are more complex, and many surface states are realized [arrows in Fig.~\ref{fig.ph_ss}(h) and (j)].

The realized phonon surface states are associated with the vibrations of the atoms at the edge of the slab.
Similar results have been reported for the two-dimensional graphene stripe~\cite{li.wang.20}.
In such a situation, the surface states separated from the bulk modes spectra~\cite{basak.ptok.23} denotes propagating of the phonon mode along the surface~\cite{herrera.kempkes.22}.
However, for the (100) surface terminated by the triangle lattice, the edge mode can be associated with a strong localized mode due to the geometry of the lattice~\cite{yang.nagaosa.14,herrera.kempkes.22}.
In the case of the ($\bar{1}$10) surface, the dispersionless edge modes along $\bar{\Gamma}$--$\bar{\text{X}}$ can be similar to the one-dimensional topological hinge electronic states protected by threefold rotational and inversion symmetries in bismuth~\cite{hsu.zhou.19}.
Similarly to the other systems, the phonon edge mode strongly depends on the surface termination~\cite{basak.ptok.23}.
Nevertheless, the topological nature of such surface states should be the source of further phonon surface states in shandite-like systems.

%%%%%%%%%%%%%%%%%%%%%%%%%%%%%%%%%%%%%%%%%%%%%
%%%%%%%%%%%%%%%%%%%%%%%%%%%%%%%%%%%%%%%%%%%%%
%%%%%%%%%%%%%%%%%%%%%%%%%%%%%%%%%%%%%%%%%%%%%

\begin{figure}
\includegraphics[width=\linewidth]{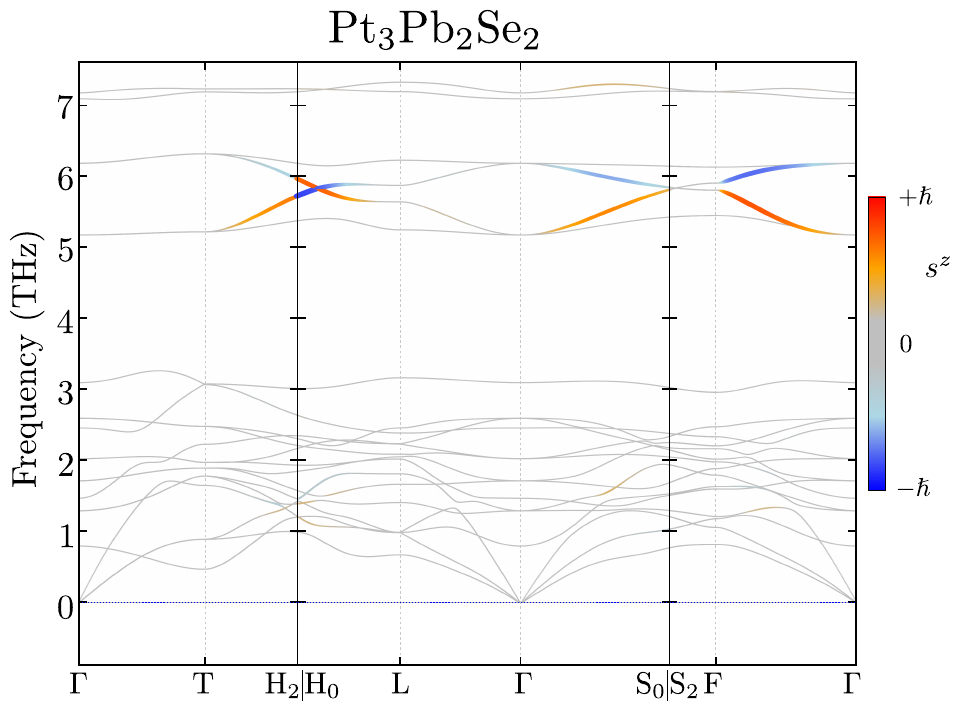}
\caption{
Chirality of the phonon modes for Pt$_{3}$Pb$_{2}$Se$_{2}$.
Color and size of the lines correspond to the phonon polarization calculated for one Pb atom.
\label{fig.chir}
}
\end{figure}

\subsection{Chiral phonons}
\label{sec.chiral}

The existence of three-fold rotational symmetry in the shandite-like structure allows for the emergence of the chiral phonons~\cite{zhang.ren.10,zhang.niu.15,liu.lian.17,chen.wu.19}, which are related to the circular motion of the atoms around the equilibrium positions.
Indeed, the analyses presented in the following show that chiral phonons can occur and are associated with Pd atoms.
However, due to the presence of inversion symmetry, the total angular momentum is equal to zero~\cite{coh.19}.
Similarly to CoSn-like compounds~\cite{ptok.kobialka.21} or layered magnetic topological insulator $T$Bi$_{2}$Te$_{4}$ ($T$=Mn,Fe)~\cite{kobialka.sternik.22}. 
Nevertheless, the translational symmetry breaking by the doping can lead to the emergence of states with non-zero total angular momentum~\cite{basak.piekarz.22,basak.ptok.22,skorka.kapcia.22}.

To study the possible realization of the phonon {\it circular polarization}, let us first define the phonon chirality.
To do this, first we should reexpress the phonon polarization vectors $\text{e}_{\bm q}$ describing vibrations of atoms induced by each phonon mode at given wavevector ${\bm k}$, in the new basis:
\mbox{$\vert R_{1} \rangle = \frac{1}{\sqrt{2}} \left( 1 \; i \; 0 \; \cdots \right)$};
\mbox{$\vert L_{1} \rangle = \frac{1}{\sqrt{2}} \left( 1 \; -i \; 0 \; \cdots \right)$};
\mbox{$\vert Z_{1} \rangle = \left( 0 \; 0 \; 1 \; \cdots \right)$};
$\cdots$;
\mbox{$\vert R_{j} \rangle = \frac{1}{\sqrt{2}} \left( \cdots \; 1 \; i \; 0 \; \cdots \right)$}; and 
\mbox{$\vert L_{j} \rangle = \frac{1}{\sqrt{2}} \left( \cdots \; 1 \; -i \; 0 \; \cdots \right)$};
\mbox{$\vert Z_{j} \rangle = \left( \cdots \; 0 \; 0 \; 1 \; \cdots \right)$}.
Here, each component of polarization vector described the motion of the atoms along $x$, $y$, and $z$ direction, while the polarization vector contains $3N$ components (where $N$ is the total number of atoms in the primitive unit cell).
In this case, we investigate the chiral modes realized in the $xy$ plane (perpendicular to the ${\bm c}$ direction). 
Our new basis denotes right-handed and left-handed circulation ($\vert R_{j} \rangle$ and $\vert L_{j} \rangle$, respectively) of the $j$-th atom.
Now, each polarization vector $\text{e}$ can be given in the form:
\begin{eqnarray}
\text{e} = \sum_{j} \left( \alpha_{j}^{R} \vert R_{j} \rangle + \alpha_{L}^{j} \vert L_{j} \rangle + \alpha_{j}^{Z} \vert Z_{j} \rangle \right) ,
\end{eqnarray}
where $\alpha_{j}^{V} = \langle V_{j} \vert \text{e} \rangle$, for~$V \in \{ R, L, Z \}$.

The operator for phonon circular polarization along the $z$-axis can be defined as:
\begin{eqnarray}
\hat{S}^{z} \equiv \sum_{j=1}^{N} s_{j}^{z} = \sum_{j=1}^{N} \left( | R_{j} \rangle \langle R_{j} | + | L_{j} \rangle \langle L_{j} | \right) , 
\end{eqnarray}
and the phonon circular polarization is equal to:
\begin{eqnarray}
s^{z}_\text{ph} = \text{e}^{\dagger} \hat{S}^{z} \text{e} = \sum_{j=1}^{N} s_{j}^{z} \hslash = \sum_{j=1}^{N} \left( | \alpha_{j}^{R} |^{2} - | \alpha_{j}^{L} |^{2} \right) \hslash ,
\end{eqnarray}
with $| s^{y}_\text{ph} | \leq 1$, since $\sum_{j} \left( | \alpha_{j}^{R} |^{2} + | \alpha_{j}^{L} |^{2} \right) = 1$.
Here, we introduce $s_{j}^{z}$, which denotes the contribution of each atom to the phonon circular polarization.
In the case of $| s_{j}^{z} | = 1$, the~$j$th atom realizes motion along an ideal circle around the equilibrium position; for~$| s_{j}^{z} | = 0$, ordinary vibrations are realized, and for $0 < | s_{j}^{z} | < 1$,  elliptic orbits are~realized.

Calculated phonon circular polarization for Pt$_{3}$Pb$_{2}$Se$_{2}$ is presented in Fig.~\ref{fig.chir}.
As we can see, at the higher frequencies branch out (from 5~THz to 6~THz) and the chiral phonon modes can be realized.
The exact analyses of the phonon polarization vectors show that the chiral modes are realized by Pb atoms.
Here, we should note that the chiral modes are realized in nondegenerate branches, coming from the decoupling of two doubly degenerate branches with $E_{\text{g}}$ and $E_{\text{u}}$ symmetry at $\Gamma$ point.

%%%%%%%%%%%%%%%%%%%%%%%%%%%%%%%%%%%%%%%%%%%%%
%%%%%%%%%%%%%%%%%%%%%%%%%%%%%%%%%%%%%%%%%%%%%
%%%%%%%%%%%%%%%%%%%%%%%%%%%%%%%%%%%%%%%%%%%%%

\begin{figure}[!t]
\includegraphics[width=\linewidth]{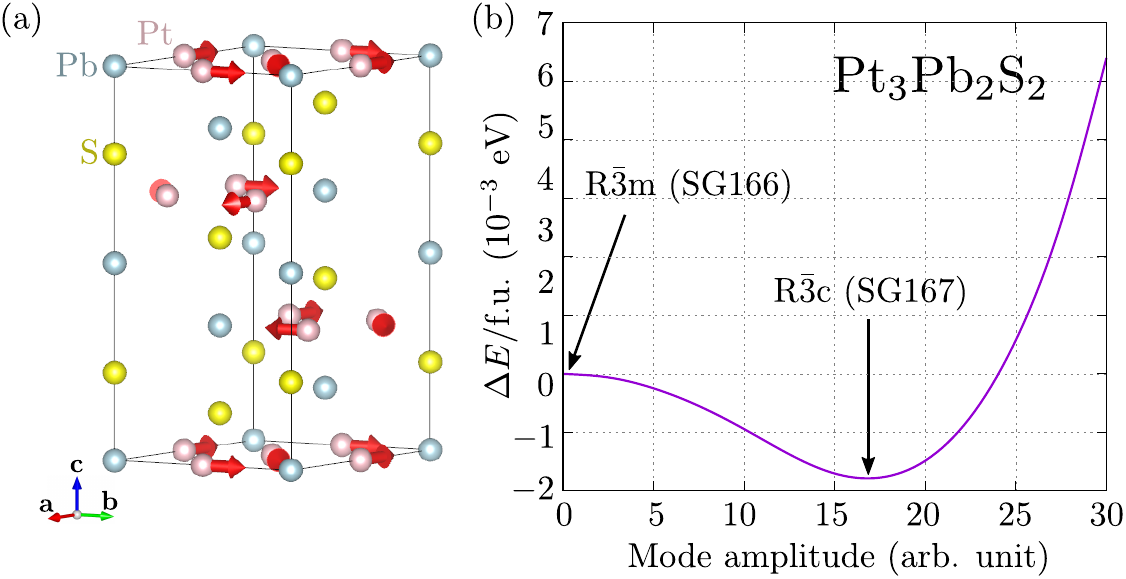}
\caption{
(a) Schematic representation of the atomic displacement pattern realized in Pt$_{3}$Pb$_{2}$S$_{2}$ with R$\bar{3}m$ symmetry.
(b) Total energy of Pt$_{3}$Pb$_{2}$S$_{2}$ as a function of the soft mode amplitude.
The relative energies of the initial R$\bar{3}$m and final R$\bar{3}$c structures are indicated by arrows.
\label{fig.ene_disp}
}
\end{figure}

\begin{figure*}
\includegraphics[width=\linewidth]{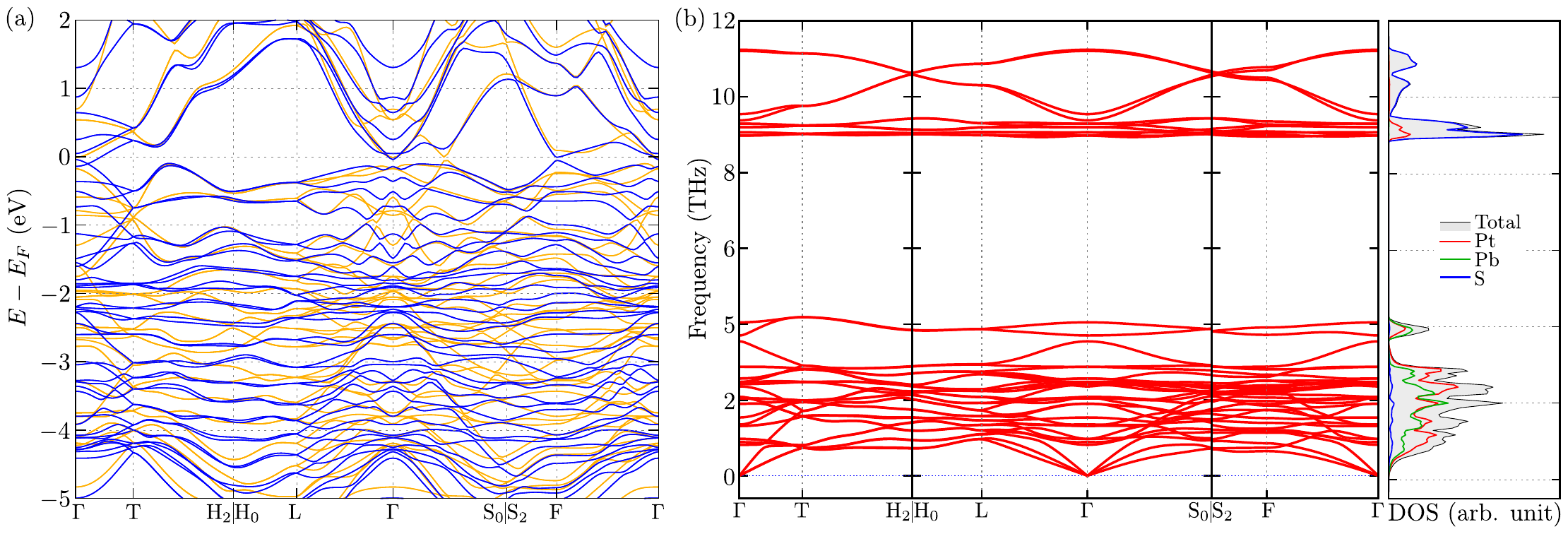}
\caption{
Total energy of Pt$_{3}$Pb$_{2}$S$_{2}$ as a function of amplitude for the soft mode at L point (left panel), and the displacement patter introduced to the crystal structure by this mode  (right panel).
Initial and displaced structures form R$\bar{3}$m and R$\bar{3}$c symmetries, respectively, with energies marked by the arrows.
\label{fig.soft}
}
\end{figure*}

\subsection{Stable Pt$_{3}$Pb$_{2}$S$_{2}$ with R$\bar{3}$c symmetry}
\label{sec.pt}

The existence of a soft mode in the Pt$_{3}$Pb$_{2}$S$_{2}$ phonon dispersion, mentioned earlier in Sec.~\ref{fig.ph_band}, indicates the existence of a collective movement of atoms that transform a higher-symmetry crystal structure into a lower-symmetry crystal structure.
Careful investigation of the atom displacement induced by the soft mode in the initial (unstable) structure can give information about the final (stable) symmetry of the system~\cite{parlinski.li.97,sternik.parlinki.05}.
In our case, the soft modes realized at $T=(1/2,1/2,1/2)$ wavevector lead not only to atom displacement, but also to an increase in the size of the primitive unit cell~\cite{sikora.gotfryd.20}, which is related to the periodicity of the induced displacement pattern.

The soft mode leads in a natural way to the symmetry breaking of the system.
Fig.~\ref{fig.ene_disp}(a) presents schematic representation of the atomic displacement pattern realized in the soft mode in Pt$_{3}$Pb$_{2}$S$_{2}$ with initial R$\bar{3}$m symmetry.
The soft mode resulted in the displacement of only the Pt atoms instead of the kagome lattice.
As a result, the symmetry breaking is related to the distortion of the kagome lattice.

The dependence of the system energy on the mode amplitude is shown in Fig.~\ref{fig.ene_disp}(b).
As we can see, the atomic displacements induced by the soft mode leads to the minimization of the energy. 
Exact system optimization leads to the system with R$\bar{3}$c symmetry (space group No.~167).
The optimized structure has lattice constants $a = 6.245$ and $c = 26.631$~\AA.
The Pt atoms are located at the Wyckoff position $18e$ (0.4812,0,1/4), the Pb atoms at two nonequivalent Wyckoff positions $6b$ (0,0,0) and $6a$ (0,0,1/2), while the S atom is located at the Wyckoff position $12c$ (0,0,0.3618).
Note that the (conventional) unit cell with R$\bar{3}$c is twice as high as the initial R$\bar{3}$m system.
In addition, by comparing both structures, we can see that the displacement of the Pt atoms is around $0.12$~\AA\ from the atomic positions in the ideal kagome sublattice.

The electronic character of Pt$_{3}$Pb$_{2}$S$_{2}$ is unchanged, and system exhibit metallic features.
The electronic band structure is more complex than for R$\bar{3}$m [Fig.~\ref{fig.soft}(a)], due to the Brillouin zone folding along (111) direction.
However, several nearly flat band are visible. 
Also, around the Fermi level, there exist a few Dirac points along $\Gamma$--T direction.
The topological properties can still be expected in this case, and thus this topic warrants futures studies in this direction.

The phonon dispersion curve does not exhibit the soft modes [Fig.~\ref{fig.soft}(b)], and Pt$_{3}$Pb$_{2}$S$_{2}$ should be stable with R$\bar{3}$c symmetry. 
In this case, the phonon modes at the $\Gamma$ point can be decomposed into irreducible representations as follows:
\begin{eqnarray}
\Gamma_\text{acoustic} &=& A_\text{2u} + E_\text{u} , \\
\nonumber \Gamma_\text{optic} &=& 2 A_\text{1g} + 3 A_\text{1u} + 4 A_\text{2g} + 4 A_\text{2u} + 7 E_\text{u} + 6 E_\text{g} .
\end{eqnarray}
The increasing number of irreducible representations is related to the doubled number of the atoms in the (primitive) unit cell.
The activity of the mode is unchanged [cf.~Eq.~(\ref{eq.irr_r3m})], i.e. modes $A_\text{2u}$ and  $E_\text{u}$ are IR active, while $A_\text{1g}$ and $E_\text{g}$ are Raman active.
The characteristic frequencies and symmetries of the modes at $\Gamma$ point for Pt$_{3}$Pb$_{2}$S$_{2}$ with R$\bar{3}$c symmetry are collected in Tab.~\ref{tab.modes2}, and can be used in the experimental classification of the real system.

\begin{table}[!t]
\caption{
\label{tab.modes2}
Characteristic frequencies (THz) and symmetries of the modes at the $\Gamma$ point for Pt$_{3}$Pb$_{2}$S$_{2}$ with R$\bar{3}$c symmetry.
}
\begin{ruledtabular}
\begin{tabular}{rlrlrlrl}
$0.82$ & ($A_\text{2u}$) & $0.88$ & ($A_\text{1u}$) & $0.97$ & ($E_\text{g}$) & $1.31$ & ($E_\text{u}$) \\
$1.35$ & ($A_\text{2g}$) & $1.54$ & ($A_\text{2g}$) & $1.90$ & ($E_\text{g}$) & $2.04$ & ($E_\text{u}$) \\
$2.08$ & ($E_\text{u}$) & $2.36$ & ($A_\text{2u}$) & $2.42$ & ($A_\text{2u}$) & $2.42$ & ($A_\text{2u}$) \\
$2.42$ & ($E_\text{u}$) & $2.42$ & ($A_\text{2u}$) & $2.48$ & ($E_\text{u}$) & $2.57$ & ($E_\text{u}$) \\
$2.57$ & ($E_\text{u}$) & $2.87$ & ($E_\text{u}$) & $3.55$ & ($A_\text{2g}$) & $3.70$ & ($A_\text{1u}$) \\
$4.05$ & ($A_\text{2u}$) & $8.98$ & ($E_\text{g}$) & $9.06$ & ($E_\text{u}$) & $9.06$ & ($E_\text{u}$) \\
$9.20$ & ($E_\text{u}$) & $9.29$ & ($E_\text{u}$) & $9.38$ & ($A_\text{2u}$) & $9.54$ & ($A_\text{2g}$) \\
$11.21$ & ($A_\text{1u}$) & $11.24$ & ($A_\text{1g}$) & & & & \\
\end{tabular}
\end{ruledtabular}
\end{table}

%%%%%%%%%%%%%%%%%%%%%%%%%%%%%%%%%%%%%%%%%%%%%
%%%%%%%%%%%%%%%%%%%%%%%%%%%%%%%%%%%%%%%%%%%%%
%%%%%%%%%%%%%%%%%%%%%%%%%%%%%%%%%%%%%%%%%%%%%

\section{Summary}
\label{sec.sum}

In this paper, we presented a comprehensive study of the dynamical properties of the $T_{3}$Pb$_{2}${\it Ch}$_{2}$ ($T$=Pd,Pt and {\it Ch}=S,Se) with shandite-like structure (R$\bar{3}$m symmetry).
We show that Pd$_{3}$Pb$_{2}${\it Ch}$_{2}$ and Pt$_{3}$Pb$_{2}$Se$_{2}$ should crystallize with shandite structure.
Contrary to this, Pt$_{3}$Pb$_{2}$S$_{2}$ is unstable dynamically, and phonon spectra exhibits the imaginary soft mode.
Analysis of this soft mode leads to the stable structure of Pt$_{3}$Pb$_{2}$S$_{2}$ with R$\bar{3}$c symmetry (without soft modes).

Furthermore, we present theoretically obtained characteristic frequencies and symmetries of the modes at the $\Gamma$ point for each compound, which should help confirm our findings experimentally, within infrared or Raman spectroscopy measurements.
From the partial phonon DOS we show that the higher frequencies vibrations are associated only with transition metal and lead atoms. 
Moreover, the branches in the higher frequencies give rise to the symmetry enforcing phonon nodal line.

The most interesting phonon properties of shandie-like compounds are associated with the emergence of the phonon surface states and chiral phonons.
We show that the phonon surface states can be realized independently of the surface termination and are mostly related to the kagome sublattice of transition metal atoms.
Existence of the three-fold rotational symmetry allows the occurrence of the chiral phonons in the system.
We show that the chiral phonons are realized by the circular motion of the Pb atoms.
Our findings can also be applicable to other compounds with the shandite R$\bar{3}$m structure.

\begin{acknowledgments}
%We warmly thank \AP{xxxx} for insightful discussions.
Some figures in this work were rendered using {\sc Vesta}~\cite{momma.izumi.11} software.
This work was supported by the Polish National Agency for Academic Exchange (NAWA, Poland) under the grant BPN/BEK/2021/1/00474 (AK).
This work was supported by the National Science Centre (NCN, Poland) under Project No. 2021/43/B/ST3/02166. %AMO
\end{acknowledgments}

\section*{Conflict of Interest}
The authors declare no conflict of interest.

\section*{Data Availability Statement}
The data that support the findings of this study are available from the corresponding author upon reasonable request.

\section*{Keywords}
Phonons, nodal lines, surface states, chiral modes

%\nocite{*}
\bibliography{biblio}

\end{document}